\documentclass[twocolumn,showpacs,preprintnumbers,amsmath,amssymb,pre]{revtex4}
\usepackage{amsmath}
\usepackage{graphicx}
\usepackage{bm}
\begin{document}
\title{Segregation Mechanisms in a Model of an Experimental Binary Granular Mixture}

\author{George C.M.A. Ehrhardt}
\altaffiliation[Now at ]{The Abdus Salam ICTP, Strada Costiera 11, 34014
Trieste (Italy)}
\email{gehrhard@ictp.trieste.it}
\author{Andrew Stephenson}
\affiliation{Theoretical Physics Group, Department of Physics and Astronomy, The University of Manchester, M13 9PL, UK.}
\author{Pedro M. Reis}
\affiliation{Manchester Centre for Nonlinear Dynamics, The University of Manchester, M13 9PL, UK.}

\date{\today}

\begin{abstract}
A simple phenomenological model of a binary granular mixture is developed and investigated numerically.  We attempt to model the experimental system of \cite{reisehrhardtstephensonandmullin,reisandmullin} where a horizontally vibrated binary monolayer was found to exhibit a transition from a mixed to a segregated state as the filling fraction of the mixture was increased.  This model is found to reproduce much of the experimentally observed behaviour, most importantly the transition from the mixed to the segregated state.  We use the model to investigate granular segregation mechanisms and explain the experimentally observed behaviour. 
\end{abstract}


\maketitle

\section{Introduction}
\label{introduction}
Granular systems exhibit a wide range of intriguing and often counterintuitive phenomena.  Segregation of two or more species of grains due to vibration or shearing is one such example \cite{mullin,ottinokhakhar}.  Many mechanisms, including buoyancy, temperature gradients, differing angles of repose, and differing roughness of particles, have been proposed \cite{ottinokhakhar,maksecizeauandstanley,cizeaumakseandstanley,shoichisim}.  
Recently, segregation of a vibrated binary mono-layer has been demonstrated in a series of experiments \cite{mullin,reisandmullin,reisehrhardtstephensonandmullin} that imply/suggest the existence of a segregation critical point as the filling fraction is varied, with associated growth of fluctuations and timescales in the vicinity of that point.

In this paper we propose a phenomenological model of the experimental system which captures its essential features.  We show that there is qualitative agreement between the model and the experiment and, in particular, that the quantitative measures reported in \cite{reisandmullin,reisehrhardtstephensonandmullin} are of the same form in both cases.  We then use the model to study segregation mechanisms, giving one definite mechanism and demonstrating that a second may also play a role.

The experimental system used in \cite{reisandmullin,reisehrhardtstephensonandmullin} consists of a smooth horizontal tray of dimensions $L_x=180mm$ by $L_y=90mm$ vibrated sinusoidally parallel to its major axis at a frequency of $f=12Hz$ and amplitude of $A=1.74\pm0.01mm$.  The grains are high-precision phosphor-bronze spheres (denoted by $c$ in the following) of radius $R_c=0.75mm$ and mass $m_c=16.8\mu g$ and poppy seeds (denoted by $p$ in the following) of average radius and mass $R_p=0.54mm$ and $m_p=0.52\mu g$ respectively.  The poppy seeds are rough, non-spherical polydisperse \cite{definitionofpolydispersity} particles.  

Particles placed on the oscillating tray predominantly move periodically with the driving.  They also have some quasi-random component to their motion, caused in the case of the $p$s by their non-sphericity which means that they do not roll smoothly but instead 'slick and slip' and scatter.  The $c$s are affected less by the driving since they roll as the tray moves beneath them.  However they too, when placed individually on the tray, are seen to have some quasi-random component to their motion.  The overall impression is of a granular bed of agitated particles with the predominant motion being parallel to the driving.

The control parameter used in the experiments was the 2-dimensional filling fraction, termed the compacity $C=(N_c \pi R_c^2 +N_p \pi R_p^2)/(L_x L_y)$, with $N_c$ fixed.  The value of $N_c$ is also important but for simplicity the experiments focussed on $N_c=1600$ and varied $N_p$.  The system was quasi 2-dimensional in that $0.49<C<1.12$, high $C$ being achieved by poppy seeds 'riding up' and overlapping each other to some extent, although never so much that they were above and overlapping the $c$s.

Starting from a uniformly mixed initial distribution, the final state reached by the system varies with $C$.  For $C \lesssim 0.65$ the system remains mixed whilst above that value small mobile segregated clusters of $c$s form.  As $C$ increases, these clusters grow in size and become anisotropic, forming stripes perpendicular to the direction of driving for $C$ large.  For $C \gtrsim 0.93$ the $c$s within these stripes crystallise to form a densely packed hexagonal lattice.  
These three phases were termed binary gas, segregation liquid, and segregation crystal.  The existence of a phase transition from the binary gas to segregation liquid, with an associated  critical value of $C$, was reported in \cite{reisandmullin}.
The details of these results are given in \cite{reisandmullin,reisehrhardtstephensonandmullin}.

The experimental system has several appealing properties from the point of view of studying granular matter: 
except at high compacities where stripes form, the final state reached is a function only of the compacity, i.e. the initial conditions are not relevant.  The particles are always in contact with the tray and hence are always effectively `thermalised'.  
This is in contrast to the behaviour of particles in granular systems such as sandpiles where the grains spend most of their time locked in position, or some vertically vibrated systems where much of the time is spent in free flight \cite{a1996reviewarticle}.
The constant 'thermalisation' of the particles, their ability to explore many possible states, and the irrelevance of initial conditions suggests that the system might be a good choice for studying the `statistical mechanics' of granular matter.  However detailed balance, equipartition of energy and other rigourous features of equilibrium statistical mechanics are of course not obeyed in this system \cite{pingpongball}.

\section{Description of Model}
\label{DescriptionofModel}

We here develop a phenomenological model of the experimental system.  The aim is to capture the essential features of the experiment in the model, thereby discovering what those features are and in particular what the segregation mechanism(s) is(are).

The following features are, we believe, necessary:
there is a tray of dimensions $L_x \times L_y$ whose base moves with sinusoidal velocity $A \sin(\omega t)$.
There are two particle species, named $c$ and $p$. 
The particles moving relative to the tray surface are subject to a frictional force.
The particles feel a randomisation of their velocity caused by their stick-slip motion with the oscillating tray and their non-sphericity (thermalisation).
The particles collide inelastically.

We therefore model the system as 2-dimensional with the particles behaving as hard disks of mass $m_{\alpha}$ and radius $R_{\alpha}$, where $\alpha$ denotes the species ($c$ or $p$).  Except during collisions the particles obey the Langevin equation
\begin{equation}
m_{\alpha} \dot{\bf v}_{\alpha i} = -\gamma_{\alpha} ({\bf v}_{\alpha i} -{\bf v}_{tray}) +{\bm \eta}_{\alpha i}(t)
\label{langevineqn}
\end{equation}
where ${\bf v}_{tray}={\bf i} A \sin(\omega t)$ and $\gamma$ provides a linear damping.  ${\bm \eta}_{\alpha}(t)$ is Gaussian white noise of mean zero and standard deviation $\left< {\bm \eta}_{\alpha}(t) \cdot {\bm \eta}_{\alpha}(t') \right> =2 \sigma^2_{\alpha} \delta(t-t')$ and this provides the `thermalisation'.  The particles interact through smooth hard-disk inelastic collisions with coefficient of restitution $r_{\alpha,\beta}$, i.e. in the centre of mass frame, ${\bf v}_{\parallel i} \to -r_{i,j} {\bf v}_{\parallel i}$, ${\bf v}_{\perp i} \to {\bf v}_{\perp i}$, where ${\bf v}_{\parallel}$ and ${\bf v}_{\perp}$ are the velocity components parallel and perpendicular to the line joining the centres of the particles $i$ and $j$.  For simplicity, the disks have been taken to be smooth sided so that angular momentum can be ignored.
A similar model has been used in \cite{colloidallanes} to describe colloidal particles driven by an external electrical field.  
Also \cite{nottsimPRE} have used a similar model without noise to model granular particles driven by a vertically oscillating air column \cite{nottscience}.

These are the essentials.  We have also kept the walls of the box stationary for simplicity but modelled the motion of the end walls by considering that in collisions they have a velocity $\max( {\bf i} A_1 \sin(\omega t) ,0)$ for the left wall and $\min( {\bf i} A_1 \sin(\omega t) ,0)$ for the right wall.  We felt that this was necessary in order to model the low-density region near each end wall caused by the vigorous collisions with the end walls.  The width of the low-density region is independent of the system size, thus we used $A_1=A/10$ rather than $A_1=A$ in the results described here in order to reduce the 'finite size effect' of this region as the system size is changed.

There are many approximations of the real system made here, the most significant ones are:
the friction term $\gamma {\bf v}$ is only an approximation, it is chosen as being the simplest possible form.
The noise is in fact due to the stick and slip interactions between the particles and the oscillatory surface, and also their non-sphericity when interacting with each other and with the tray.  We do not try to directly model this since it would require detailed specification of the shape of each particle and its actual interaction with the tray, which is not known.  Even a single high-precision phosphor-bronze sphere conducts a quasi-random walk when placed on the oscillating tray, indicating that the randomness can depend on very small imperfections of the particles and the tray (and possibly also in the driving).  Both because of this immense difficulty and in order to have a reasonably simple model whose behaviour we can understand, we instead choose to include the noise phenomenologically.  We assume that the noise the particles receive is independent of their neighbours and of the phase of the tray cycle, both of which are unlikely to be accurately what happens in the experiments.  The assumption that the noise is Gaussian and white is an approximation.
For simplicity we are using a 2-dimensional model, which ignores the overlapping of $p$s and the rolling of particles.  The particles and walls are assumed to be smooth and so angular momentum is not considered.
We have ignored the polydispersity of the $p$s.  Polydispersity was included to check its importance and was found to leave the qualitative behaviour unchanged.
The final approximation is that the coefficients of restitution are constant which is a commonly made one \cite{somegenersimulationref}. 
Despite these simplifications and approximations, in Section \ref{Results} we show that our phenomenological model captures much of the behaviour observed experimentally.

\subsection{Parameter Values}
\label{ParameterValues}

Static parameter values such as mass and size can be measured reasonably accurately, for the poppy seeds we have used the mean values of a sample of measurements \cite{howmany}.  Dynamic parameters were less accurately known, $\gamma_{\alpha}$ was estimated from the distances travelled by single particles striking the moving end walls.  Using the result for the noiseless case,
\begin{equation}
x(\infty)=x(0)+v_x(0) m/\gamma
\label{}
\end{equation}
and estimating $v_x(0)$ to be equal to the maximum velocity of the end wall gives approximate values for $\gamma_{\alpha}$.  The value for the noise is the hardest to determine since no velocity or accurate ${\bf r}(t)$ path measurements were available.  We merely estimated that the mean square velocity due to the noise, $\left< {{\bf v}_{\alpha}}^2 \right> = \sigma_{\alpha}^2 m_{\alpha}/\gamma_{\alpha}$ should be equal to $\approx (A/F)^2$ where the factor $F=3$ for the $p$ case and $F=13$ for the $c$ case.

Clearly these last estimates are rather crude.  However, extensive study of a wide range of these parameters has shown that the qualitative features are 
robust to variation of these estimates. 

The coefficients of restitution, $r_{\alpha,\beta}=r_{\beta,\alpha,}$ are estimated to be: $r_{c,c}=0.9$, $r_{c,p}=0.2$, $r_{p,p}=0.1$, $r_{c,w}=0.9$, $r_{p,w}=0.2$ where $w$ denotes a side wall.  Other values have been studied, but the qualitative behaviour is unchanged.   The main effect of increasing(decreasing) $r_{\alpha,\beta}$ is to increase(decrease) the granular temperature which in general merely moves the onset of segregation to slightly higher(lower) compacity values.

The parameters then are the following.
\begin{center}
\begin{tabular}{||l||l|l||} \hline 
 property  &   $c$ value & $p$ value    
\\  \hline  
$m_{\alpha}$ mass                 & $1.6800 \times 10^{-5}$   & $5.2000 \times 10^{-7}$ \\ 
$R_{\alpha}$ radius               & $7.5000 \times 10^{-4}$   & $5.4000 \times 10^{-4}$ \\ 
$\gamma_{\alpha}$ damping term    & $4.3636 \times 10^{-6}$   & $1.0000 \times 10^{-5}$ \\ 
${\sigma_x}_{\alpha}$ noise term  & $8.1819 \times 10^{-8}$   & $1.0000 \times 10^{-7}$ \\ 
${\sigma_y}_{\alpha}$ noise term  & $8.1819 \times 10^{-8}$   & $1.0000 \times 10^{-7}$ \\ 
$r_{\alpha,\beta}$                & $r_{c,c}=r_{c,w}=0.9$     & $r_{c,p}=r_{p,w}=0.2$   \\
$r_{\alpha,\beta}$                & $r_{c,p}=0.2$             &  $r_{p,p}=0.1$          \\
\hline
\end{tabular} 
\end{center}
\label{parameterstable}
\noindent Table \ref{parameterstable}.  The parameters used for the simulation results described below (SI units are used at all times unless stated otherwise).  ${\sigma_x}_{\alpha}$ and ${\sigma_y}_{\alpha}$ are the $x$ and $y$ components of ${\bf \sigma}$.
\medskip

\subsection{Simulation Method}
\label{SimulationMethod}

We have simulated the model via an event driven code \cite{somegenersimulationref}.  The process is as follows:
\begin{enumerate}
\setlength{\parskip}{-3pt}
\item For each particle, predict when it will next collide.
\item Identify the first collision to occur.
\item Move the particle(s) involved so that they touch.
\item Update the velocities of the particle(s) involved (change in velocity due to damping and noise since they were last updated).
\item Collide the particle(s) \cite{infrequentcase}.
\item Re-predict the next collision(s) of the particle(s) and their neighbours.
\item Repeat from 2 until time has advanced by $t_{minupdate}/2$ 
\item Update all particles that have not been updated in the last $t_{minupdate}/2$ seconds.
\item Repeat from 1 until time has advanced by $t_{takedata}$.
\item Record data.
\item Repeat from 1.
\end{enumerate}

The prediction of collisions assumes that the particle's velocities do not change during their motion.  The error caused by this is small provided that $t_{minupdate} \ll \min(\tau_{\alpha})$ where $\tau_{\alpha}=m_{\alpha}/\gamma_{\alpha}$ is the time constant of the velocity decay.  We set $t_{minupdate}=0.01 \times \min(\tau_{\alpha})$.

The hard sphere model with inelastic collisions can undergo inelastic collapse \cite{inelasticcollapseref}, where the particles undergo an infinite number of collisions in a finite time.  Clearly a simulation that implements each collision will `stall' in these circumstances.  One way around this unphysical singularity is the tc model \cite{tcmodel} which prevents the collapse by setting $r_{\alpha,\beta}=1$ for any particle which collided within the last $t_{tcolmin}$ seconds.  We found that for $t_{tcolmin}<1 \times 10^{-4}$ 
the results did not change.  For the results presented here we chose $t_{tcolmin}=5 \times 10^{-5}$. 

The initial conditions were created by running a reduced size system $N_c/25$, $N_p/25$, of twice the aspect ratio with particles initially placed randomly on a square lattice.  An external force was applied to compress the particles into an area of size $L_x/5 \times L_y/5$.  The particles were then allowed to move in this box without the external force until equilibrated.  The initial condition was then created by tessellating the full system size with 25 replicas of the reduced system.  To prevent long-range order the replicas were randomly inverted in both the $x$ and $y$ directions (the differing noises received by the particles in conjunction with the chaotic behaviour of the particles would rapidly remove any correlations in any case).  During this stage, both particle species had the same properties, including $r_{\alpha,\beta}=1$, except for their radii.
This method produced initial conditions which appeared to be as homogeneous as those for the experiment.

The simulations were run on standard PCs.  For comparison with the experiments our results are for the same system size ($18cm \times 9cm$), except where stated otherwise.  The aspect ratio is here $2:1$ in all cases. 

\section{Results}
\label{Results}

As with the experiments, we used the number of $p$s, $N_p$ as our control parameter.  The main quantities measured were those found experimentally \cite{mullin,reisandmullin,reisehrhardtstephensonandmullin} and related to the $c$s only: the mean stripe width (MSW), and the local density, $\rho_i$, of the $i$th particle.  We also visualised the system and watched its behaviour.  In addition, we also measured the area available to the $p$s, and the kinetic energy or 'granular temperature' of the particles. 

\begin{figure} \begin{center}
\includegraphics[width=0.5\columnwidth]{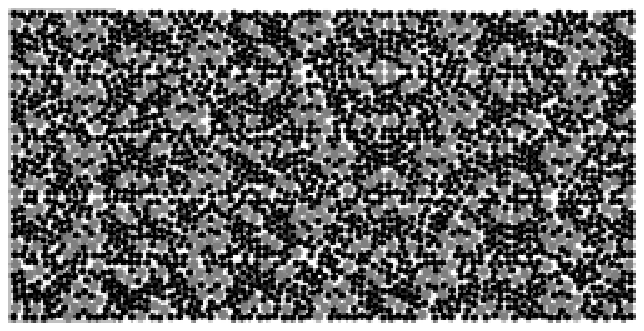}\includegraphics[width=0.5\columnwidth]{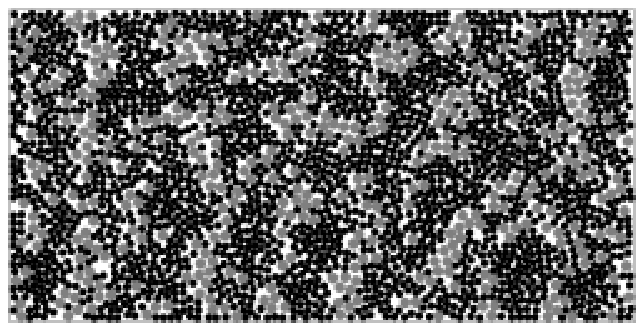} 
\includegraphics[width=0.5\columnwidth]{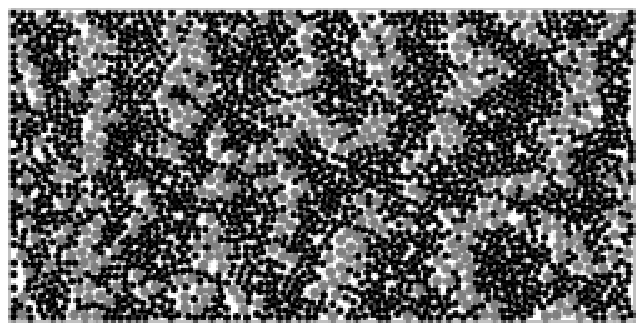}\includegraphics[width=0.5\columnwidth]{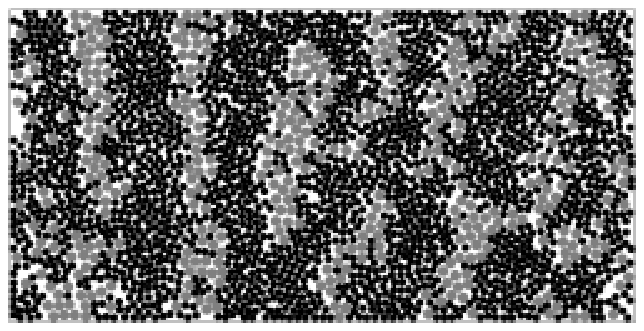} 
\includegraphics[width=0.5\columnwidth]{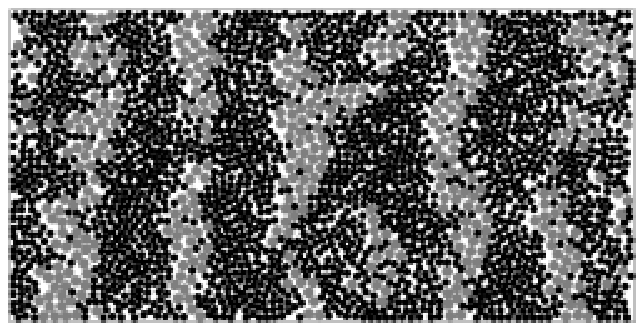}\includegraphics[width=0.5\columnwidth]{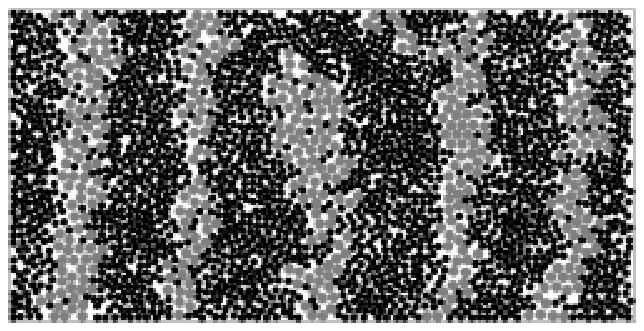} 
\caption{The time evolution of a simulation showing the coarsening
  into domains for a system of $1/4$ the area of the experiment and
  $C=0.6721$.  The times are, top left: t=0.04s, top right: t=4.18s,
  middle left: t=8.37s, middle right: t=16.75s, bottom left: t=33.51s,
  bottom right: t=62.83s.  The $p$s are coloured black. \label{bob}}
\end{center} \end{figure}

The MSW measures the mean width of the stripes in the $x$ direction by pixellating an image of the system, deciding which pixels $P_{x,y}$ are within a $c$ domain, then running along each pixel row $P_y$ and counting the width and number of domains.  All the rows $P_y$ are summed and the average domain width found.   Whether $P_{x,y}$ is within a domain is decided by blurring the image with a gaussian smoothing function and then setting a threshold.  This measure was used in \cite{reisandmullin,reisehrhardtstephensonandmullin} even when the domains had not formed into stripes since it provided a simple measure of the domain sizes in the longitudinal direction.

The normalised local density $\rho_i$ was found by Voronoi tessellation \cite{voronoiref} of the $c$s, such that each $c_i$ has around it a polygonal area all points of which are closer to $c_i$ than to any other $c$.  $\rho_i$ is then the minimum possible area, which is $2 \sqrt{3} R_c^2$, divided by the polygonal area.  Polygonal cells on the edge of the system which are not bounded are discarded.

In addition to the MSW as a measure of the amount of coarsening, we also measured the area available to the $p$s.  This 'available area' is just the fractional area of the system in which a $p$ could be placed without overlapping a $c$.  Each $c$ has a circular 'excluded area' around it of radius $Rc+Rp$, inside which the centre of a $p$ cannot be placed.  If all $N_c$ $c$s are widely separated, the available area is $1-N_c \,\, \pi (Rc+Rp)^2/Lx Ly$, whilst if all the $c$s are hexagonally packed in one domain the available area will be larger ($\approx 1-N_c \,\, 2 \sqrt{3} Rc^2/Lx Ly$) since the excluded areas now overlap.  Thus the available area gives a measure of how segregated the $c$s are.  In Section \ref{NoiseSegregation} we will show that the available area is relevant to noise segregation.

\begin{figure} \begin{center}
\includegraphics[width=0.9\columnwidth]{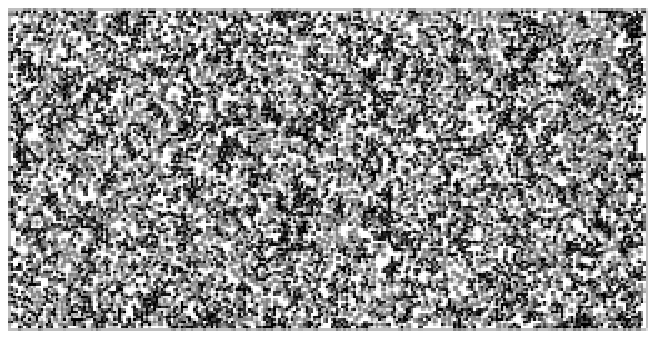} 
\includegraphics[width=0.9\columnwidth]{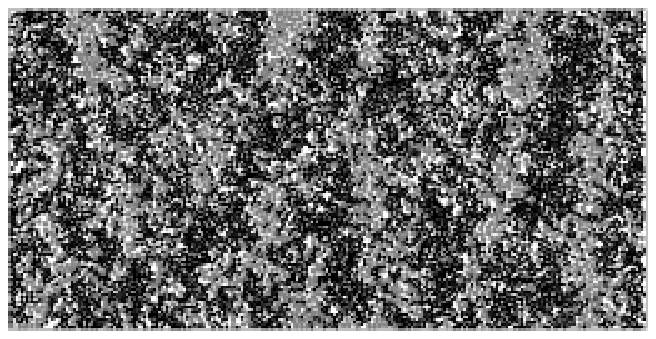} 
\includegraphics[width=0.9\columnwidth]{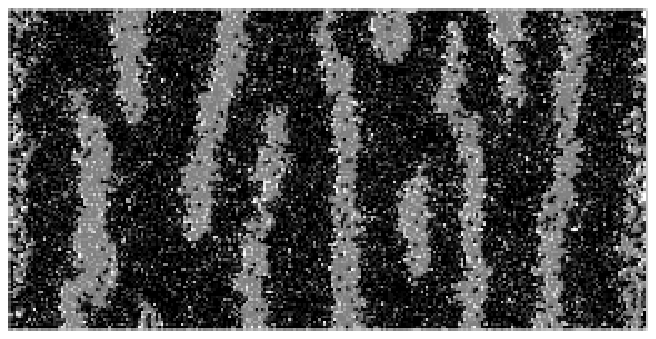} 
\caption{Examples of a binary mixture, segregated liquid, and segregated stripes. Compacities are 0.446, 
0.582, 
and 0.717, 
respectively.  $p$s are coloured black.  The pictures were taken after 100 seconds for the gas and 200 seconds for the liquid and crystal, by which time the first two have reached a steady state whilst the third is still slowly evolving.  \label{snapshotsofsystem}}
\end{center} \end{figure}

\begin{figure} \begin{center}
\includegraphics[width=0.9\columnwidth]{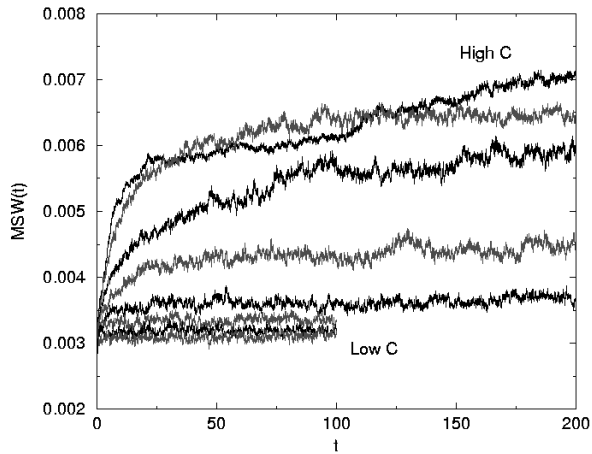} 
\caption{Plots of MSW against time for various compacities.  From the right, top to bottom, the compacities decrease from $C$=0.717 
to $C$=0.401 
in uniform increments.  Thus it can be seen that for low $C$ the system does not coarsen whilst when $C$ is increased, the system coarsens to a roughly constant value which increases with $C$.  For the largest value of $C$, the system has undergone rapid initial coarsening but then slowed as the large domains move more slowly, especially at this high compacity where the system is becoming somewhat `jammed'.    
The MSW was measured every 0.0209 
seconds.  \label{mswsvst}}
\end{center} \end{figure}

\begin{figure} \begin{center}
\includegraphics[width=0.9\columnwidth]{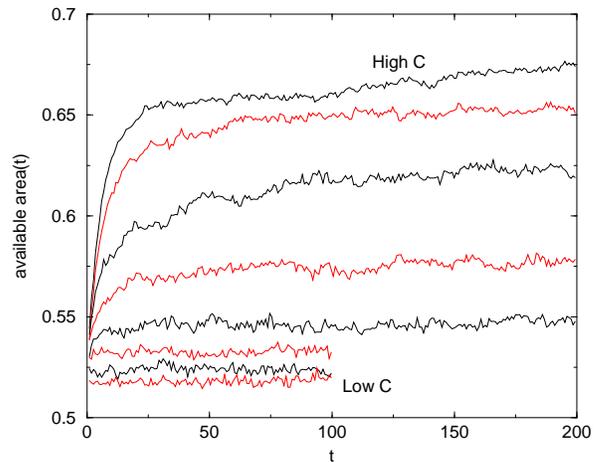} 
\caption{Plots of the available volume for the $p$s against time for various compacities.  From the right, top to bottom, the compacities decrease from $C$=0.717 
to $C$=0.401 
in uniform increments.  The conclusions are similar to those for figure \ref{mswsvst} except that the curves here always remain in the same order.  The available volume was measured every 0.838 
seconds.  \label{availablevolumevst}}
\end{center} \end{figure}

Figure \ref{bob} shows 6 images of the evolution of a coarsening system.  It can be seen that the initial segregation into relatively small domains is rapid.  This is followed by slower coarsening as domains merge and as larger domains grow at the expense of smaller ones that 'evaporate'.  This growth rate is also clearly seen in the time-series plots of figures \ref{mswsvst} and \ref{availablevolumevst}.
Figure \ref{snapshotsofsystem} shows three images of the system at differing compacities for late times, the lowest compacity ($C$=0.446) 
 shows a binary mixture, the second ($C$=0.582) 
a segregated `liquid' which has mobile and transient clusters and shows a slight anisotropy, and the third ($C$=0.717) 
shows a system that has coarsened into stripes.  The first two are in a steady state whilst the third is still evolving slowly due to the high compacity (for a strictly 2-dimensional system) which causes particles to be `blocked' by other particles.

Figure \ref{mswsvst} shows MSW$_C$(t) for several compacities.  As $C$ increases, the $c$s, which were initially mixed with the $p$s, coarsen into domains whose size increases with $C$.  The early-time coarsening is rapid, followed by slower coarsening and then saturation at some relatively steady value.  For the lowest compacities, there is no coarsening, the system remains in a mixed, disordered state.  The highest $C$ value shown has not reached a final steady state, the stripes are still moving and merging at a very slow rate compared to the initial coarsening.  At higher compacities, the system becomes blocked or jammed, the particles being unable to rearrange themselves in 2-dimensions, the slow timescales of the top curve show the onset of this jamming.  This jamming does not occur to the same extent in the experimental system since the particles do not always form a monolayer and $p$s can move out of the way of $c$s by 'riding up' on top of each other.

Figure \ref{availablevolumevst} shows a plot of the value of the area available to the $p$s against time, it is similar to that for the MSW.  It is a more repeatable measure than the MSW since it is less affected by stripes merging, for example, two runs with the same parameters naturally differ due to the chaotic behaviour of the particles.  This means that for high compacities merging of stripes in two realizations at the same parameters may occur at different times, causing the $MSW_C(t)$ curves to differ between runs at late times (since $MSW_C(t)$ is inversely proportional to the number of domains).  The available volume is less affected by this and thus provides a 'cleaner' measure of the coarsening.  Thus for later results we will use the available volume, although the MSW provides similar, if more noisy, results.  
Experimentally, the available volume has not been measured since the experimental system was only imaged in it's central region, thus the number of $c$s changes with time making the measure somewhat arbitrary (the changes due to $c$s entering or leaving the system would be more significant than the coarsening).
We note that for the highest compacity curve, the area available has increased by $\approx 0.15$ whilst the maximum range of the available area stated before is $\approx 0.30$.

\begin{figure} \begin{center}
\includegraphics[width=0.75\columnwidth,angle=270]{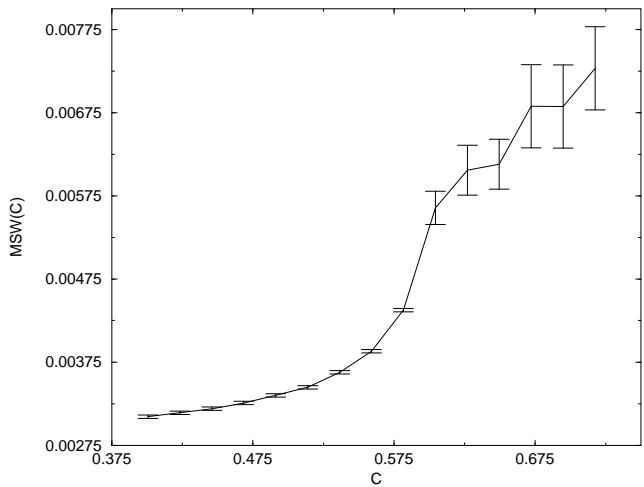} 
\caption{ Plot of the saturated (late time) mean stripe width as a function of compacity.  The saturated values were found by fitting exponentials to the MSW(t) curves.  Two runs were done at each compacity, and the results averaged.  The error bars are based on the difference between these two runs. 
\label{mswvsc}}
\end{center} \end{figure}

\begin{figure} \begin{center}
\includegraphics[width=0.8\columnwidth,angle=270]{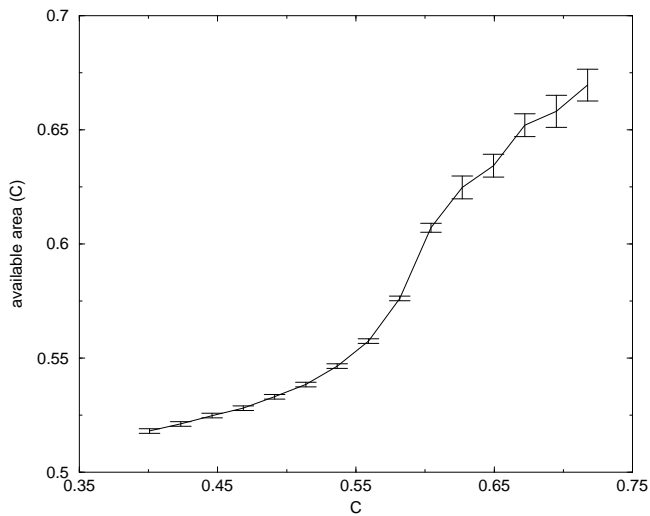} 
\caption{ Plot of the saturated (late time) area available to the $p$s
  as a function of compacity.  The saturated values were found by
  fitting exponentials to the available-area(t) curves.  Two runs were
  done at each compacity, and the results averaged.  The error bars
  are based on the difference between these two runs.
Note that the error bars are smaller than for the MSW vs $C$ plot since, as stated in the text, the available area is a 'cleaner' measure.
\label{exvsc}}
\end{center} \end{figure}

In figures \ref{mswvsc} and \ref{exvsc} we plot the late time values of the MSW and available volume as a function of the compacity.  This was done by fitting exponentials to the timeseries of the type shown in figures \ref{mswsvst} and \ref{availablevolumevst} and using the late-time values.  We do not claim that the time series are exponential, but the fits provide us with a reasonable measure of the late-time values.  
The MSW vs $C$ curve is the same measure as that used experimentally in \cite{reisandmullin,capri} to claim a mixed state to segregated state phase transition for the experimental system.  In \cite{reisandmullin} a square root curve was fitted to the right hand side of the data, the data to the left of the transition being taken to be roughly constant, i.e. $MSW_{sat}(C)= {\it B}$ for $C<C_{transition}$ and  $MSW_{sat}(C) = {\it B} + {\it D} \sqrt{C-C_{transition}}$ for $C>C_{transition}$, where ${\it B}$ and ${\it D}$ are constants.  From figure \ref{mswvsc} we see that this is not the case for the simulation data, which, as $C$ is increased, initially rises slowly, then  increasingly rapidly before slowing again for large $C$.  There is no indication of a discontinuity in the gradient of the order parameter, merely a rapid increase.  In \cite{capri} the same experimental data is presented with a sigmoid shaped 'guide to the eye' curve rather than a square root.  In our opinion the simulation data is similar to the experimental data but not to the square root form suggested in \cite{reisandmullin}. 
The main conclusion from both the experimental and simulation $MSW_{sat}(C)$ curves is that $MSW_{sat}(C)$ has a roughly sigmoid shape, with $MSW_{sat}(C)$ increasing rapidly with $C$ in the central region.

The simulation value of $C$ at which the rapid increase occurs ($\approx 0.58$) differs somewhat from the experimental value ($0.647 \pm 0.049$) \cite{reisandmullin}.  In the next section we show how this value changes continuously as we vary the noise strength or other parameters.  Thus, by increasing the noise strength we can increase the value of $C$ at which the rapid increase occurs.

\begin{figure} \begin{center}
\includegraphics[width=0.85\columnwidth,angle=270]{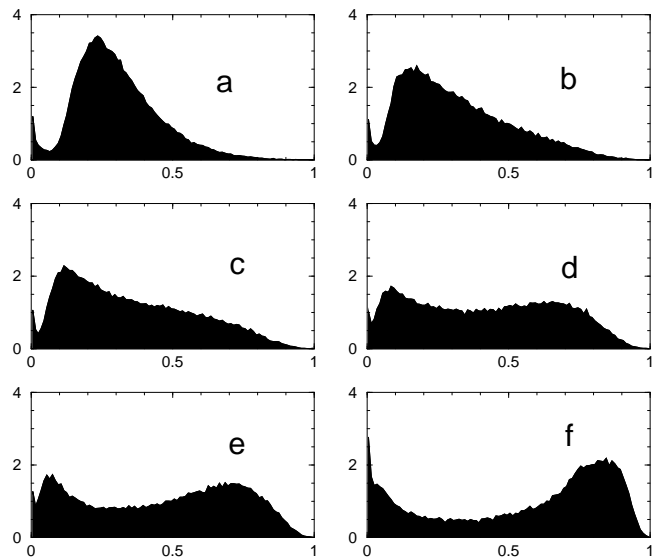} 
\caption{ Normalised histograms of the Voronoi densities plotted against normalised density, for $C$=0.400728(a),
0.559(b), 
0.582(c), 
0.604(d), 
0.627(e), 
and 0.740(f). 
Note the crossover with increasing $C$ from a single peak on the left to two peaks, then a larger peak on the right.   For each compacity the data was measured in the steady state from 45 frames of 1600 $c$s each.
\label{voronoihistograms}}
\end{center} \end{figure}

\begin{figure} \begin{center}
\includegraphics[width=0.8\columnwidth,angle=270]{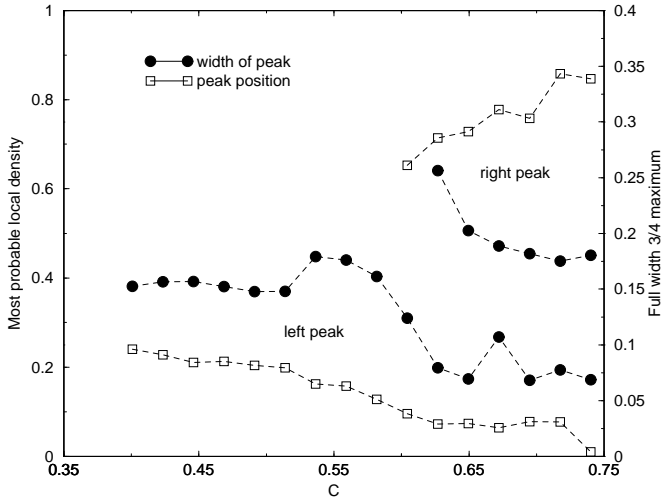} 
\caption{ Plots of data derived from the Voronoi histograms as shown in figure \ref{voronoihistograms}.  Squares show the position of the peak(s) in the histograms, representing the 'most probable' densities, the lower curve is for the leftmost peak and the upper curve is for the rightmost peak.  Filled circles show the full width of each peak at $3/4$ maximum (where the peak does have a distinct $3/4$ maximum).  The curves differ from their experimental equivalents where there was only one maximum (thus just two single curves) which moved steadily to the right and where the width was peaked at an intermediate compacity.  These differences are due mainly to the presence in the simulation density histograms of a second peak at low density which is always present and is caused by a small number of isolated $c$s, something not observed in the experiment.
\label{voronoihistogramsderiveddata}}
\end{center} \end{figure}

Figure \ref{voronoihistograms} shows histograms of the local Voronoi density for various compacities, the results are qualitatively similar to the experimentally reported ones \cite{reisehrhardtstephensonandmullin}.  At the lowest $C$ values, corresponding to the mixed state, the distribution is peaked at low densities as one would expect for unclustered $c$s.  At high $C$ the distribution is peaked at large densities as one would expect for clustered (i.e. segregated) $c$s.  
There is a crossover between these two cases, with figure \ref{voronoihistograms}d showing a broad histogram due to almost the full range of densities being present in almost equal weights. 
Unlike the experimental results however, the high $C$ distribution also has a peak at low density caused by a small fraction of isolated $c$s which are not present in the experiment.  There is a crossover between these two extremes as $C$ is increased, the central 4 figures (b-e) show this crossover.  
Following \cite{reisehrhardtstephensonandmullin}, we plot the location of the peak(s) and their widths as a function of compacity in figure \ref{voronoihistogramsderiveddata}.  In \cite{reisehrhardtstephensonandmullin} the peak width used was the full width $3/4$ maximum since the peak did not extend far enough above the rest of the distribution for a full width $1/2$ maximum to be meaningful.  Here we did the same although a full width $3/4$ maximum also does not exist in one case.  Figure \ref{voronoihistogramsderiveddata} is qualitatively different from its experimental equivalent and we conclude that although the histograms are qualitatively similar, the results derived from them are not.

The results presented here show that our model reproduces the segregation and its qualitative behaviour with $C$ of binary gas, segregated liquid, and segregated stripes as seen experimentally.  This behaviour is not immediately obvious from the model rules - it emerged from the set of rules which we believed contained the important microscopic features of the experiment.
We have also shown that the MSW as a function of time and its saturated value as a function of compacity behave in a qualitatively similar manner to the experiment.
These reproductions of experimentally observed behaviour lead us to conclude that the `necessary features' listed in section \ref{DescriptionofModel} capture the essential behaviour of the system.  
Nonetheless, our phenomenological model does not quantitatively reproduce the experiment (it was never expected to do so), in particular the Voronoi histograms differ from the experimental ones sufficiently that the data derived from them (figure \ref{voronoihistogramsderiveddata}) is qualitatively different.

\section{Segregation Mechanisms}
\label{SegregationMechanisms}
Having shown that our model is relevant to the experiment, we now use it to investigate segregation mechanisms.

\subsection{Segregation due to Oscillatory Driving}
\label{OscillatoryDriving}

That the domains are anisotropic and indeed form stripes at high $C$ indicates that the anisotropy of the driving is significant.  Experimentally the stripes form perpendicular to the driving even for aspect ratios greater than $1$, e.g. $L_y/L_x=2$.

The side-to-side driving causes the two species to move at different rates due to their differing masses and friction coefficients.  A single particle will oscillate with $\left< x_{\alpha}(t) \right>= A / \sqrt{1+\omega^2 (m_{\alpha} / \gamma_{\alpha})^2} \,\, \sin(\omega t)$ where the average is over the noise.  Thus $\left< x_c(t) \right>=6.00\times 10^{-6} \sin(\omega t)$ and $\left< x_p(t) \right>=4.30 \times 10^{-4} \sin(\omega t)$. 
Thus the $p$s would 'like' to move a distance of order their radius during a cycle whilst the $c$s hardly move.  Consider a state of only $p$s, if we remove the noise then all the $p$s move in the same sinusoidal way, if we transfer to the (non-inertial) reference frame in which they are at rest, we see that this is identical to the state with no sinusoidal driving (apart from edge effects at the walls) for which the dissipation causes the particles to be stationary.  The same can be said of a state of only $c$s.  In a mixed state, however, the $p$s will collide with the $c$s and the system will `scatter' into a different state.  Stable states, i.e. those that do not undergo further scatterings, will be those for which the $c$s are separated from the $p$s in the $x$ direction by distances of at least the amplitude of the $p$s' oscillations.  For the packing fractions considered here, this can only happen by the two species segregating into domains.  The fact that the area available to the $p$s increases with time is consistent with this interpretation.  This argument only holds when the driving is not too large compared to the dissipation.  For example if the amplitude of oscillation were of order the system size then it would not be valid.  An argument similar to this was also given in \cite{nottsimPRE}.

We have studied systems with very low noise and the results obtained agree with the heuristic argument given above.  The addition of noise, which causes particles to diffuse, will tend to cause mixing.  Thus there is a competition between the periodic driving which causes segregation and the noise which prevents it.  This is shown in the results presented below.

\begin{figure} \begin{center}
\includegraphics[width=0.7\columnwidth,angle=270]{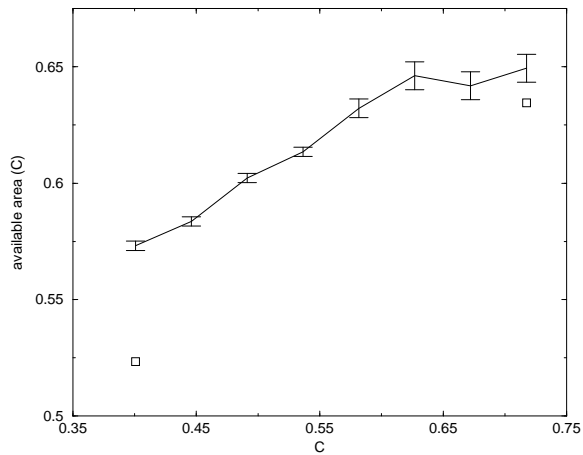} 
\caption{ Plot of the area available to the $p$s at late times as a function of compacity.  The curve shows standard parameters but with low noise ($\sigma_{\alpha} \to 0.1*\sigma_{\alpha}$) results, the two box points are the equivalent results for the standard parameters.  Thus it can be seen that all compacities have segregated.  For high $C$ a few large domains form whilst at lower compacities there is enough space for a larger number of small domains to be stable.    Note that the area of the system is $1/4$ that of the experimental system.  The late-time results were found by fitting an exponential to the available area vs time curves.
\label{standardparametersbutlownoise}}
\end{center} \end{figure}

\begin{figure} \begin{center}
\includegraphics[width=0.7\columnwidth,angle=270]{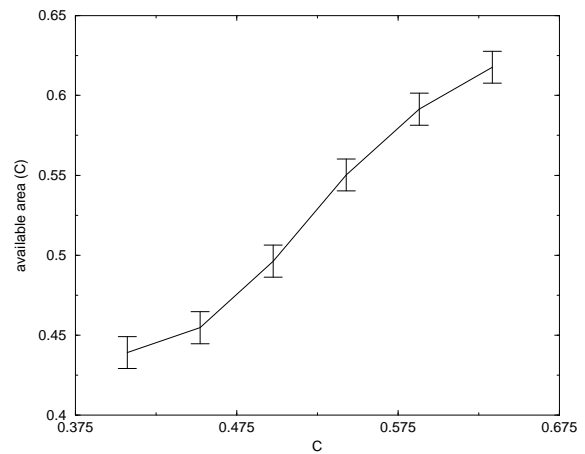} 
\caption{ Plot of the area available to the $p$s at late times as a function of compacity.  The data is for a system where both particle types are the same (standard $p$ parameters) except that only the $p$s feel the periodic driving.  We get a mixed state for low $C$ and a segregated state for high $C$, as we did for the standard parameters.  Note that the area of the system is $1/4$ that of the experimental system.  The late-time results were found by fitting an exponential to the available area vs time curves.
\label{allsameexceptris1andonlypsfeelsidetosideSEGREGATEorMIX}}
\end{center} \end{figure}

For the standard parameters but with $\sigma_{\alpha} \to 0.1 \times \sigma_{\alpha}$, the system segregates for all the compacities studied in section \ref{Results} (see figure \ref{standardparametersbutlownoise}), indicating that, as expected, the noise acts to prevent segregation.  
To confirm this we then gave the two species identical parameters (the $p$ parameters) and set all $r_{\alpha,\beta}=1$.  The only difference was that the $p$s experienced the periodic driving term $\gamma {\bf v}_{tray}$ whilst the $c$s did not.  The results are shown in figure \ref{allsameexceptris1andonlypsfeelsidetosideSEGREGATEorMIX} and demonstrate that a difference in the periodic driving alone can cause segregation.

\begin{figure} \begin{center}
\includegraphics[width=0.7\columnwidth,angle=270]{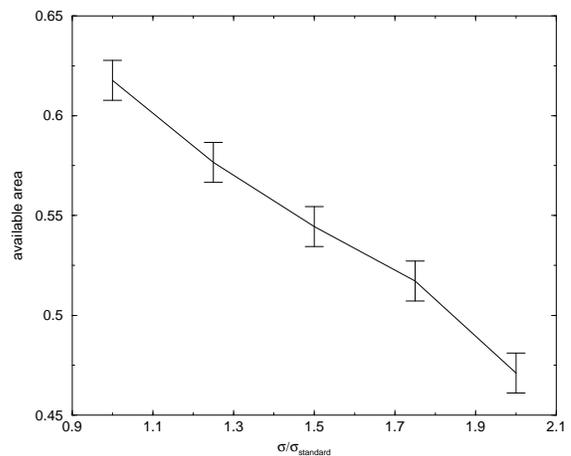} 
\caption{ Plot of the area available to the $p$s at late times as a function of the noise strength $\sigma/\sigma_{standard}$, for $C=0.498$. 
Increasing the noise strength brings us from a segregated to a mixed state.  Note that the area of the system is $1/4$ that of the experimental system.  The late-time results were found by fitting an exponential to the available area vs time curves.  \label{allsameexceptris1andonlypsfeelsidetosideSEGREGATEorMIXsigmascan}}
\end{center} \end{figure}

For $C=0.498$ 
with the same parameters except $\gamma_{\alpha} \to 2 \times \gamma_{\alpha}$ (to reduce the granular `temperature' which is higher than usual due to the $r=1$) we varied $\sigma$ and, as shown in figure \ref{allsameexceptris1andonlypsfeelsidetosideSEGREGATEorMIXsigmascan}, the system segregates for low $\sigma$ and remains mixed for high $\sigma$.

\begin{figure} \begin{center}
\includegraphics[width=0.7\columnwidth,angle=270]{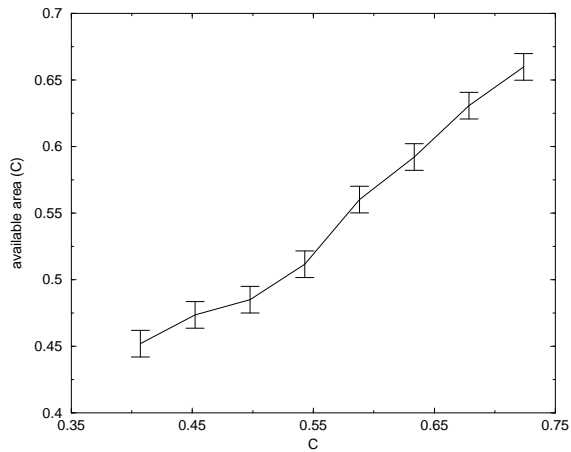} 
\caption{ Plot of the area available to the $p$s as a function of compacity, the parameters are as for figure \ref{allsameexceptris1andonlypsfeelsidetosideSEGREGATEorMIX} except that here the coefficients of restitution are all $0.1$ rather than $1$.  It can be seen that this does not qualitatively change the results.  Note that the area of the system is $1/4$ that of the experimental system.  
\label{allsameexceptris0pt1andonlypsfeelsidetoside}}
\end{center} \end{figure}

These results confirm that the state of the system is a result of a competition between the periodic driving which causes segregation and the noise or `granular temperature' which prevents it.  Varying our control parameter, $N_p$, causes us to go from a mixed to a segregated state because increasing $N_p$ both reduces the granular temperature (since it increases the number of collisions which are highly inelastic) and also increases the `pressure' that the $c$s feel due to collisions with a greater number of oscillating $p$s.  Figure \ref{allsameexceptris0pt1andonlypsfeelsidetoside} shows results for all properties set to $p$ values including all $r=0.1$ but with only the $p$s feeling the periodic driving, showing that the results remain of the same form for $r < 1$.

Finally, we find that the system with standard parameters does not segregate if the oscillatory motion is turned off.

These results show that the differential oscillatory driving can cause segregation and present good evidence that it is the responsible mechanism in the system in conjunction with the noise which acts to prevent segregation.  The transition from a mixed to a segregated state as $C$ is varied is due to the compacity changing the relative strengths of these two competing effects.

\subsection{Noise Segregation}
\label{NoiseSegregation}

It was suggested in \cite{reisandmullin,reisehrhardtstephensonandmullin} that the segregation mechanism might be similar to the depletion interaction in equilibrium binary systems \cite{entropicorderingrefs}.  As an example consider a colloidal suspension containing non-adsorbing polymers.  In the ideal case where there are no forces present, the free energy depends only on the entropy of the system.  Treating the polymers as spheres \cite{altenburgnotesref}, it is clear that each colloidal particle has an `excluded volume' around it of radius $R_{colloid} +R_{sphere}$ which the centre of the polymer cannot enter.  Thus the volume available to the polymers is the volume of the system less the excluded volumes around the colloidal particles and the system edges.  However, the excluded volumes overlap when colloidal particles are closer than $2 (R_{colloid} +R_{sphere})$, thus the volume available to the polymers, and hence their entropy, is larger if the colloidal particles are close to each other.  This entropic `effective potential' can be large enough to cause the colloid to coagulate.  This mechanism was one reason for measuring the area available to the $p$s in our simulations.
Segregation has been observed in simulations of hard spheres of two different sizes \cite{whysegregationsohardtoobserve} for large size ratios, e.g. $R_1/R_2=10$, and also experimentally in binary mixtures of hard sphere colloids \cite{1995dinsmore}.  

We may equivalently view the entropy argument from the kinetic point of view.  Two particles which are close to each other such that no third particle may fit in the space between them will feel a pressure on all sides due to collisions with other particles $\it{except}$ on their neighbouring sides.  Thus the particles feel an effective attractive force.  This pressure argument may be extended to systems that are not in equilibrium, for example the granular experiment studied here.  The size ratio $R_c/R_p$ is much closer to unity than for simulated equilibrium segregating systems \cite{whysegregationsohardtoobserve}, implying that the difference in size alone does not cause segregation.  

\begin{figure} \begin{center}
\includegraphics[width=0.9\columnwidth]{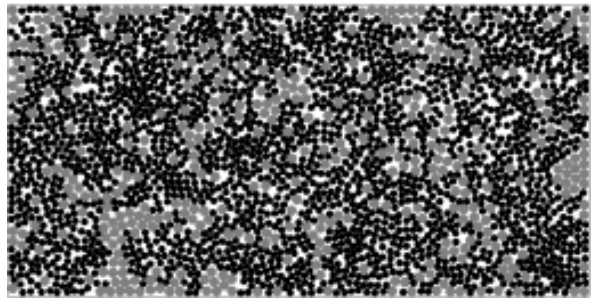}
\caption{Segregation of $c$s for the standard parameters but with no
  periodic driving and $\sigma_c \to 0.05 \times \sigma_c$.  Notice
  the domains of $c$s at the edges of the system.  Note that the area
  of the system is $1/4$ that of the experimental system.  $p$s are coloured black.   
\label{figanexampleoftrueparamstimes0pt05segregation}} 
\end{center} \end{figure}

\begin{figure} \begin{center}
\includegraphics[width=0.9\columnwidth]{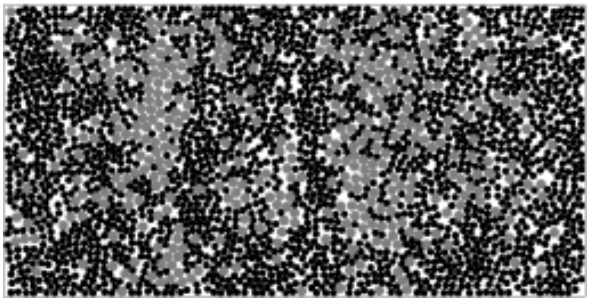}
\caption{Segregation of $c$s for standard parameters but with
  $\sigma_{c\,x} \to 0.05\times\sigma_{c\,x}$ and $\sigma_{p\,x} \to
  2\times\sigma_{p\,x}$ and no driving of the tray base.  Thus we get segregation but without coagulation at the top and bottom walls.  The left and right walls are oscillating and prevent coagulation there.  Note that the area of the system is $1/4$ that of the experimental system.  $p$s are coloured black. \label{figanexampleofpressuresegregationwithoutgroupingatwalls}} 
\end{center} \end{figure}

In our out-of-equilibrium system there are several  possible differences between the two species besides a difference in size.  For our system, it seemed possible that the lighter, faster moving $p$s might, through the differential pressure mechanism, cause the $c$s to coagulate even in the absence of periodic driving.  As stated before, this was not observed for the standard parameters.  We therefore increased the noise of the $p$s and/or reduced the noise of the $c$s in order to increase the $p$ to $c$ temperature and hence pressure ratio.
Noticeable segregation occurred over a wide range of compacity values for $\sigma_p \to 10 \times \sigma_p$, for $\sigma_p \to 2 \times \sigma_p$ and $\sigma_c \to 0.1 \times \sigma_c$, and also for $\sigma_c \to 0.05 \times \sigma_c$.  Whilst the first of these cases is outside the reasonable parameter range, the second and third are at parameters which might be physical.  Figure \ref{figanexampleoftrueparamstimes0pt05segregation} shows an example for the third case.  Notice that many of the $c$s have coagulated at the walls as one would expect since the pressure argument for two-particle attraction also applies to the particle-wall case.  Although this marked congregation at the walls is observed in experiments with colloids, it is not observed in the experiments of \cite{reisandmullin,reisehrhardtstephensonandmullin}.  Re-introducing the periodic driving of the end walls prevented congregation at the $x=0,L_x$ walls whose large momentum transfer to the particles, as stated earlier, gives rise to low-density regions next to them. However, for all parameter values studied that displayed segregation with no driving, the $c$s still congregated at the $y=0,L_y$ walls when the driving was turned on.  Since stripes that touch the top and bottom walls are stable in the experiment, it seems unlikely that agitation due to the motion of the top and bottom walls is what prevents the liquid domains from coagulating there.
It is possible to remove this experimentally unobserved effect by giving $\sigma_x$ and $\sigma_y$ differing values such that $\sigma_{xc}$ is lower and $\sigma_{yc}$ is higher than that needed to produce segregation.  Whilst we had previously kept $\sigma_x=\sigma_y$ for simplicity, it is reasonable that $\sigma_x > \sigma_y$ since random motion caused by sticking and slipping is likely to be larger in the direction of driving.  This extra modification produces segregation without $c$s congregating on the walls (provided that the $x=0,L_x$ walls are `driving') whether there is periodic driving of the tray base or not.  Figure \ref{figanexampleofpressuresegregationwithoutgroupingatwalls} shows an example.

We therefore conclude that this differential pressure segregation mechanism may play a role in the experiment.  We had to `tune' the parameters which implies that the mechanism is less robust than the oscillatory driving mechanism.  Accurate experimental measurements of the parameters would help to resolve whether this mechanism is indeed present.  Directly distinguishing the two mechanisms mentioned would require accurate tracking of all particles and their collisions, coupled with investigations of other binary mixtures in order to get readings for different noise to side-to-side movement ratios.  This is likely to prove a difficult task and at present all we can conclude is that differential pressure segregation may play a role in the experiment in addition to the differential driving discussed above.

\begin{figure} \begin{center}
\includegraphics[width=0.9\columnwidth]{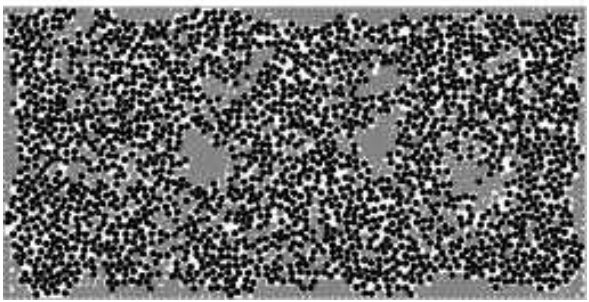}
\caption{Coagulation of $c$s after 96 seconds due only to temperature difference Tp=1600*Tc.  The system is still evolving, the groups of $c$s in the centre will eventually attach to the sides of the system.  Note that the area of the system is $1/4$ that of the experimental system.  $p$s are coloured black. \label{figdifferenttscausecoarsening}} 
\end{center} \end{figure}

To demonstrate that different temperatures alone can cause segregation we have studied two species with standard $p$ parameters but all $r_{\alpha,\beta}=1$.  The temperature difference is produced by $\sigma_p \to 40 \times \sigma$ and also setting all $\gamma \to 100 \times \gamma$ so that the time constants $\tau = m / \gamma$ are sufficiently small that particles remember the temperature of their heat baths rather than only the temperature of their previous collision partners.  This imposed temperature difference causes coagulation of the lower temperature particles as shown in figure \ref{figdifferenttscausecoarsening}.  Whilst these parameters are very different from the standard ones, they clearly show another segregation mechanism in a nonequilibrium system and that the heuristic arguments regarding pressure differences out of equilibrium are valid.

\subsection{Further Mechanisms}
\label{FurtherMechanisms}

One further possible mechanism is that the noise of the particles is correlated among the particles but differently for the $c$s and the $p$s.  This could cause segregation in the same way as the periodic driving since it also would produce different collective motions for the two species.  The correlation could be caused by, for example, the $p$s all changing from sticking to the tray base to slipping at the same time in the periodic cycle.  Whilst this mechanism is at least plausible, the collective motions due to correlated noise and due to periodic driving (which would have caused the correlated noise in the first place) would not be clearly distinguishable.  At the level of this phenomenological model which breaks the driving into a periodic and a noise component, any such mechanism is, therefore, not meaningful.

It is possible that there are other segregation mechanisms not discussed here, however we believe that we have considered the ones most likely to be relevant in the experiment.  

\section{Conclusion}
\label{Conclusion}

In this paper we have introduced and numerically studied a (relatively) simple phenomenological model of a recently reported granular segregation experiment.  We have measured the same quantities as measured experimentally and shown that our model reproduces most of the features of the experiment, the most important being a transition from a mixed to a segregated state as the compacity is increased.  This behaviour is not \emph{a priori} built into the model - it emerges from the simple rules governing the motion of the particles.  This is significant as it shows that we have a set of basic features necessary for an explanation of the experimentally observed behaviour.

We then used our model to investigate and identify segregation mechanisms and elucidate the experimental behaviour.  
We showed that the transition from mixed to segregated state in the model is caused by competition between the different driving felt by the $c$s and $p$s, which acts to cause segregation, and the noise, which acts to prevent segregation.  We are led to conclude that this is also the main mechanism present in the experiment.
We have also considered and demonstrated segregation due to different pressures and shown that it is possible that this might play a role in the experiment.
The differential driving segregation mechanism is applicable to many binary driven systems \cite{colloidallanes,nottscience}.

This work goes some way to explaining the intriguing experimental results of \cite{mullin,reisandmullin,reisehrhardtstephensonandmullin}.  Experiments with other particle types in conjunction with more accurate experimental measurements, particularly with regard to particle positions and velocities, should allow more accurate comparisons with our model and also refinements of the model.  In particular more accurate values for the parameters used.  

Now that we have shown our model to be relevant to the experimental system, it is possible to use it further to investigate the 'granular statistical mechanics' of this type of system.  In particular, it may be of use in developing and testing theories for agitated granular mixtures before attempting the more difficult task of accurately comparing with experiments.

\begin{acknowledgments}
G.E. thanks Timo Aspelmeier and Sam Carr for useful discussions.  P.M.R. thanks Tom Mullin for many stimulating conversations.  P.M.R. was supported by a doctoral scholarship from the Portuguese Foundation of Science and Technology.
\end{acknowledgments}


\begin{thebibliography}{99}
\bibitem{howmany} 2500 $p$s were measured to calculate the mean radius.
\bibitem{infrequentcase}  In the uncommon event that the particles are no longer moving together after the velocity update, they are not collided.
\end{thebibliography}
\end{document}